 \def\Title{The Schr{\"o}dinger-HJW Theorem}
 \def\arXiv{quant-ph/0305068v3}
 \def\Abstract{%
A concise presentation of Schr{\"o}dinger's ancilla theorem 
(1936 \emph{Proc.~Camb.~Phil.~Soc.} \textbf{32}, 446) and its several recent rediscoveries.
}%
\documentclass[aps,rmp,nobibnotes,nofootinbib,onecolumn,twoside,noshowpacs,draft]{revtex4}
\makeatletter 
\def\frontmatter@setup{\normalfont\raggedright}
\makeatother
 \usepackage[tbtags,sumlimits]{amsmath}
 \usepackage{amssymb}
 \makeatletter
 \def\p@section{}
 \def\p@subsection{}
 \def\p@subsubsection{}
 \def\p@paragraph{}
 \def\p@subparagraph{}

 \def\@hangfrom@section#1#2#3{\@hangfrom{#1}{\large\textrm{#2}}{\large\textrm{#3}}}%
 \def\@hangfrom@subsection#1#2#3{\@hangfrom{#1}{\textrm{#2}}{\textrm{#3}}}%
 \def\@hangfrom@subsubsection#1#2#3{\@hangfrom{#1}{\textrm{#2}}{\textrm{#3}}}%
 \def\frontmatter@setup{\normalfont\raggedright}
 \makeatother

 \usepackage{xspace}

 \newcommand{\ie}{i.e., }
 \newcommand{\eg}{e.g., }

 \newcommand{\RefEqn}[1]{Eq.~\eqref{#1}}

 \newcommand{\QED}{\hfill\ensuremath{\square}\smallskip\smallskip}
 \newcommand{\Proof}{\par\noindent\emph{Proof:~}}
 \newcommand{\IFF}{\/i\/f\/f\/\xspace}

 \newcommand{\set}[1]{\ensuremath{{\left\{\,#1\,\right\}}}}%
 \newcommand{\setsuch}[2]{\ensuremath{\big\{\,#1\,\big|\,#2\,\big\}}}%

 \newcommand{\abs}[1]{\ensuremath{\left\vert#1\right\vert}}
 
 \newcommand{\KDelta}[2]{\ensuremath{\delta_{{#1}{#2}}}}

 \newcommand{\Op}[1]{\ensuremath{\boldsymbol{#1}}\xspace}
 \newcommand{\One}{\ensuremath{\mathbf{\displaystyle{1}}}\xspace}

 \newcommand{\Sys}[1]{\ensuremath{\mathcal{#1}}\xspace}

 \newcommand{\ket}[1]{\ensuremath{\vert\,{#1}\,\rangle}}
 \newcommand{\bra}[1]{\ensuremath{\langle\,#1\,\vert}}
 \newcommand{\braket}[2]{\ensuremath{\langle#1\,\vert\,#2\rangle}}
 \newcommand{\proj}[1]{\ensuremath{\ket{{#1}}\bra{{#1}}}}

 \newcommand{\PSI}[2]{\Psi^{\Sys{#1}}_{\,#2}}%
 \newcommand{\ketPsi}[2]{\ensuremath{\left|\PSI{#1}{#2}\right\rangle}\xspace}

 \newcommand{\bRho}{\rho}
 \newcommand{\Rho}[2]{\ensuremath{\bRho^{\Sys{#1}}_{#2}}\xspace}

 \newcommand{\Trace}[2][]{\ensuremath{{\rm Tr}_{\Sys{#1}}\left\{\,{#2}\right\}}}

 \newcommand{\HS}[1]{\ensuremath{\mathcal{H}{}^{\Sys{#1}}}\xspace}

 \newcommand{\Schrodinger}{Schr\"odinger\xspace}
 \newcommand{\OneS}{\One^{\SysS}}
 \newcommand{\utr}{\ensuremath{\boldsymbol{\rm{U}}}\xspace}

 \newcommand{\SysS}{\Sys{S}}
 \newcommand{\SysM}{\Sys{M}}
 \newcommand{\SysSM}{\Sys{S\oplus M}}
 \newcommand{\HSS}{\HS{S}}
 \newcommand{\HSM}{\HS{M}}
 \newcommand{\HSX}{\HS{X}}
 \newcommand{\HSSM}{\ensuremath{\HSS\otimes\HSM}}
 \newcommand{\HSSMX}{\ensuremath{\HSS\otimes\HSM\otimes\HSX}}
 \newcommand{\RhoS}[1][]{\Rho{S}{#1}}
 
 \newcommand{\RhoSM}[1][]{\Rho{S\oplus M}{#1}}
 \newcommand{\ketPsiSM}{\ketPsi{SM}{}}
 \newcommand{\U}[1][]{\Op{{\rm U}^{\Sys{#1}}}}

 \newcommand{\kPSM}{\ensuremath{|\Psi^{\mathcal{SM}}\rangle}\xspace} 

 \newcommand{\nS}{\ensuremath{n_{\SysS}}\xspace}
 \newcommand{\nM}{\ensuremath{n_{\SysM}}\xspace}

 \newcommand{\RhoEnsemble}{{$\bRho$}-ensemble\xspace}
 \newcommand{\RhoSensemble}{{\RhoS}\!-ensemble\xspace}
\begin{document}
 \makeatletter
 \def\ps@titlepage{%
   \renewcommand{\@oddfoot}{}%
   \renewcommand{\@evenfoot}{}%
   \renewcommand{\@oddhead}{\emph{Found.~Phys.~Lett.} {\bf19}(1) 95-102 (2006)\hfill\arXiv}}
 \makeatother
 \flushbottom
\title[Kirkpatrick -- \Title]{\Title}
 \author{K.~A.~Kirkpatrick}
 \affiliation{New Mexico Highlands University, Las Vegas, New Mexico 87701} 
 \email[E-mail: ]{kirkpatrick@physics.nmhu.edu}
\begin{abstract}
 \Abstract
\end{abstract}
 \maketitle
 \makeatletter\markboth{\hfill\@shorttitle\hfill}{\hfill\@shorttitle\hfill}\makeatother
 \pagestyle{myheadings}

\vspace{-2em}

\section{Introduction}
We re-present a theorem which \Schrodinger \nocite{Schrodinger36} proved in 1936. He commented that this theorem was one ``for which I claim no priority but the permission of deducing it in the following section, for it is certainly not \emph{well} known.'' His comment was amusingly prescient: The theorem was rediscovered by \citeauthor{Jaynes57} in 1957 (whose work was extended by \citet{Hadjisavvas81}), rediscovered by Hughston, Jozsa, and Wootters (HJW) in \citeyear{HughstonJW93} (this last an expansion of a 1989 partial rediscovery by \citeauthor{Gisin89}); in 1999, \citeauthor*{Mermin99} simplified a portion of HJW's proof --- and it would appear none of these were aware of \Schrodinger's work. Furthering the irony, Mermin commented that this is ``a pertinent theorem which deserves to be more widely known.''

But not only more widely known; this theorem deserves treatment in terms of physically relevant ancillae (following Mermin) rather than formal transformations by orthonormal-column matrices, and it deserves a complete statement and a concise proof in one place. Here is our attempt at such a presentation.

\section{Preliminaries}
\renewcommand{\baselinestretch}{1.1}\tiny\normalsize
Throughout, \HSS and \HSM are Hilbert spaces with dimensions \nS and \nM, respectively, \Trace[M]{\cdot} is the trace over \HSM of an operator on \HSSM, \kPSM is a vector in \HSSM, and $\RhoS=\Trace[M]{\proj{\Psi^{\Sys{SM}}}\,}$. The dimension of the support of \RhoS is $n_{\rho}$, $n_{\rho}\leq\nS$.

\par\smallskip\noindent\textbf{Lemma.~}
\emph{%
\ket{\chi} and \ket{\phi} are vectors in \HSSM. If $\Trace[M]{\proj{\chi}}=\Trace[M]{\proj{\phi}}$, then there exists a unitary transformation \U on \HSM such that $\ket{\chi}=\big(\OneS\otimes\U\big)\ket{\phi}$. 
}\smallskip%

\Proof  The operator $\Op{\rm X}\equiv\Trace[M]{\proj{\chi}}$ is positive and Hermitian, so its eigenvalues are non-negative and its eigenkets $\set{\ket{p_j}}_{j=1}^{\nS}$ are an orthonormal basis of \HSS; thus 
\begin{equation}\label{E:X}
\Op{\rm X}=\sum_{s=1}^n\,w_s\,\proj{p_s},\quad\text{with}\quad 
  \begin{aligned}
    &w_j>0,\; 1\leq j\leq n\leq\nS,\\
    &w_j=0,\;j>n.
  \end{aligned}
\end{equation}
For \set{\ket{\mu_j}} any basis of \HSM, $\ket{\chi}=\sum_{s=1}^{n_{\SysS}}\sum_{t=1}^{n_{\SysM}}\psi_{st}\,\ket{p_s\,\mu_t}$; setting $\ket{\eta_s}=\sum_{t=1}^{n_{\SysM}}\psi_{st}\,\ket{\mu_t}$, we have $\ket{\chi}=\sum_{s=1}^{n_{\SysS}}\ket{p_s\eta_s}$. Then
\begin{equation}
 \Op{\rm X}=\sum_{s}^{\nS}\ket{p_s}\bra{p_{s'}}\braket{\eta_{s'}}{\eta_s}=
 \sum_{s=1}^n\,w_s\,\proj{p_s},
\end{equation}
whence $\braket{\eta_{s'}}{\eta_s}=w_s\KDelta{s}{s'}$. Hence the $\set{\ket{b_j}\equiv\ket{\eta_j}/\sqrt{w_j}}_{j=1}^{n}$ are orthonormal and we may write $\ket{\chi}=\sum_{s=1}^n\,\sqrt{w_s}\,\ket{p_s\,b_s}$. A similar argument leads to $\ket{\phi}=\sum_{j=1}^n\,\sqrt{w_j}\,\ket{p_j\,c_j}$, the \set{\ket{c_j}} orthonormal. Extend the sets \set{\ket{b_j}} and \set{\ket{c_j}} to orthonormal bases of \HSM and set $\U=\sum_{s=1}^{\nM}\,\ket{b_s}\bra{c_s}$; clearly, \U is unitary and $\OneS\otimes\U$ performs the desired transformation. \QED

\par\noindent\textbf{Definition~}(\textsl{{\RhoSensemble}}).
Given a positive, Hermitian, unit trace operator \RhoS, a \emph{\RhoSensemble of order $n$ ($n\geq n_{\rho}$}) is a set $\set{\big(\ket{\phi_j}\in\HSS,\,w_j>0\big)}_{j=1}^n$, with $\sum_{s=1}^n w_s=1$ and the $\set{\ket{\phi_j}}_{j=1}^n$ noncollinear, such that $\RhoS=\sum_{s=1}^n w_s\proj{\phi_s}$. We will call a \RhoSensemble \emph{linearly independent} if the $\set{\ket{\phi_j}}_{j=1}^n$ are linearly independent. 

\par\smallskip\noindent\textbf{Definition~}(\textsl{Ancilla}).
A set $\set{\ket{b_j}\in\HSM}_{j=1}^n$ is an \emph{ancilla} of the \RhoSensemble $\set{\big(\ket{\phi_j}\in\HSS,\,w_j>0\big)}_{j=1}^n$ \IFF it is an orthonormal set and $\ketPsiSM=\sum_{s=1}^{n}\phi_s\ket{\phi_s\,b_s}$, $\abs{\phi_j}^2=w_j$.

\par\smallskip\noindent\textbf{Definition~}(\textsl{U-map}).
The \RhoSensemble $\set{\big(\ket{\psi_k},\,v_k\big)}_{k=1}^m$ is \emph{U-mapped} to the \RhoSensemble $\set{\big(\ket{\phi_j},\,w_j\big)}_{j=1}^n$ \IFF there exists a set \begin{equation}
  \setsuch{u_{jk}\in \mathcal{C}}{\sum_{t=1}^n u_{tk}u^*_{tk'}=
    \KDelta{k}{k'}}_{(j,k)=(1,1)}^{(n,m)}
\end{equation}
such that 
\begin{equation}
  \left\{\sqrt{w_j}\,\ket{\phi_j}=
    \sum_{t=1}^m\,u_{jt}\sqrt{v_t}\,\ket{\psi_t}\right\}_{j=1}^n.
\end{equation} 
A unitary transformation \utr on \HSM \emph{generates} this U-map \IFF $\set{u_{jk}=\bra{b_j}\utr\ket{b_k}}_{(j,k)=(1,1)}^{(n,m)}$, with $\set{\ket{b_j}\in\HSM}_{j=1}^{\max(n,m)}$ the ancilla of the greater-order \RhoSensemble.

\section{The Theorem}
\medskip\par\noindent\textbf{Theorem~}(\Schrodinger-HJW). %
{\itshape
The state of $\mathcal{S\!\oplus\!M}$ is \kPSM; $\RhoS\!=\!\Trace[M]{\proj{\!\Psi^{\Sys{SM}}}}$. Then
\par\noindent(a). Every \RhoSensemble of order $n\leq n_{\SysM}$ has a corresponding ancilla in \HSM. If the \RhoSensemble is linearly independent, the ancilla is unique. 
\par\noindent(b). Every orthonormal basis of \HSM contains exactly one ancilla corresponding to exactly one \RhoSensemble. 
\par\noindent(c). Given any two {\RhoSensemble}s, there exist unitary transformations on \HSM which generate a U-map from one to the other; if each \RhoSensemble is linearly independent, the transformation is unique. 
\par\noindent(d). Every unitary transformation on \HSM generates a U-map of every \RhoSensemble to another.
\par\noindent(e). Every vector in the support of \RhoS appears as an element of at least one \RhoSensemble.
}
\smallskip%

\Proof 
\par\noindent(a). Given the \RhoSensemble $\set{\big(\ket{\phi_j},\,w_j\big)}_{j=1}^n$, use an arbitrary orthonormal set $\set{\ket{d_j}\in\HSM}_{j=1}^n$ to construct the vector $\ket{\Psi'}=\sum_{s=1}^{n}\phi_s\ket{\phi_s\,d_s}$. By the Lemma, there exists a unitary transform \U on \HSM such that
\begin{equation}
 \ketPsiSM=\big(\OneS\otimes\U\big)\ket{\Psi'}=%
 \sum_{s=1}^{n}\phi_s\ket{\phi_s}\otimes\U\,\ket{d_s};
\end{equation}
the ancilla is the orthonormal set $\set{\U\,\ket{d_j}}_{j=1}^n$. Clearly, if the \set{\ket{\phi_j}} are linearly independent, that ancilla is unique.

\smallskip\noindent(b). \set{\ket{b_k}} is an arbitrary orthonormal basis of \HSM; expand \ketPsiSM in terms of it and any basis \set{\ket{p_j}} of \HSS: $\ketPsiSM=\sum_{j=1}^{n_{\SysS}}\sum_{k=1}^{n_{\SysM}}\alpha_{jk}\ket{p_j\,b_k}$. Define $\phi_k\ket{\phi_k}\equiv\sum_{j=1}^{n_{\SysS}}\alpha_{jk}\ket{p_j}$ and $w_k\equiv\abs{\phi_k}^2$, and re-order the index $k$ so $w_k>0,\,k\leq n\leq n_{\SysM},\;w_k=0,\,k>n$; then 
\begin{equation}
\ketPsiSM=\sum_{j=1}^{n}\phi_j\ket{\phi_j\,b_j}.
\end{equation}
Thus $\set{\ket{b_j}}_{j=1}^n$ is an ancilla, and $\set{\big(\ket{\phi_j},\,w_j\big)}_{j=1}^n$ is the corresponding \RhoSensemble. 
Suppose the set $\set{\ket{b_j}}_{j=1}^{n_{\SysM}}$ contained two ancillae, say $\setsuch{\ket{b_j}}{j\in\mathcal{A}}$ and $\setsuch{\ket{b_k}}{k\in\mathcal{B}}$, with $\mathcal{A}, \mathcal{B}\subset \set{1\dots n_{\SysM}}$:
\begin{equation}
 \ketPsiSM=\sum_{s\in\mathcal{A}}\phi_s\ket{\phi_s\,b_s}=%
            \sum_{t\in\mathcal{B}}\psi_t\ket{\psi_t\,b_t}.
\end{equation}
\noindent The orthonormality of the \set{\ket{b_j}} guarantees $\mathcal{A}=\mathcal{B}$, hence the uniqueness of the ancilla within \set{\ket{b_j}}, and it guarantees $\ket{\phi_j}=\ket{\psi_j}$ and $\phi_j=\psi_j$ for all $j$ in $\mathcal{A}$, hence the uniqueness of the \RhoSensemble corresponding to that ancilla.

\smallskip\noindent(c). Given two {\RhoSensemble}s $\set{\big(\ket{\phi_j},\,w_j\big)}_{j=1}^m$ and $\set{\big(\ket{\psi_k},\,v_k\big)}_{k=1}^n$, (a) guarantees ancillae $\set{\ket{b_j}\in\HSM}_{j=1}^m$ and $\set{\ket{c_k}\in\HSM}_{k=1}^n$, respectively, with
\begin{equation}\label{E:both}
\ketPsiSM=\sum_{s=1}^m\sqrt{w_s}\,\ket{\phi_s\,b_s}=\sum_{t=1}^n\sqrt{v_t}\,\ket{\psi_t\,c_t}
\end{equation}
(phases are absorbed into the ancillary kets). Extend \set{\ket{b_j}} and \set{\ket{c_k}} to orthonormal bases of \HSM; the unitary operator $\utr=\sum_{t=1}^{n_{\SysM}}\ket{c_t}\bra{b_t}$ transforms $\ket{c_k}=\utr\,\ket{b_k}=\sum_{s=1}^{n_{\SysM}}u_{sk}\ket{b_s}$, where $u_{jk}=\bra{b_j}\,\utr\,\ket{b_k}=\braket{b_j}{c_k}$, so
\begin{equation}\label{E:either}
 \ketPsiSM=%
\sum_{s=1}^{n_{\SysM}}\bigg(\sum_{t=1}^{n}\sqrt{v_t}\,u_{st}\,\ket{\psi_t}\bigg)\ket{b_s};
\end{equation}
Equating \eqref{E:both} and \eqref{E:either}, we obtain the \utr-generated U-map
\begin{equation}\label{E:mapping}
  \sum_{t=1}^{n}\sqrt{v_t}\,u_{jt}\,\ket{\psi_t}=
    \begin{cases}
       \sqrt{w_j}\,\ket{\phi_j}&j\in\set{1\cdots m}\\
       0&\text{otherwise}.
    \end{cases}
\end{equation}
If each \RhoSensemble is linearly independent, each ancilla is unique, so the transformation between them is unique (modulo element-label permutation).

\smallskip\noindent(d). Given any \RhoSensemble, (a) guarantees an ancilla; this may be extended to an orthonormal basis of \HSM. Any unitary transformation maps this to another orthonormal basis; by~(b), this second basis contains a single ancilla correlated to a single \RhoSensemble. By \RefEqn{E:mapping}, this unitary transformation provides the U-mapping between these two {\RhoSensemble}s.

\smallskip\noindent(e). \ket{\xi} is an arbitrary vector in the support of \RhoS; we will construct a \RhoSensemble with \ket{\xi} as its first element. The Schmidt expression of \ketPsiSM, $\sum_{s=1}^n\psi_s\ket{p_s\,a_s}$, gives a basis for the expansion our arbitrary vector: $\ket{\xi}=\sum_{s=1}^n\gamma_s\ket{p_s}$, with $\sum_{s=1}^n|\gamma_s|^2=1$. For any $\nM\times\nM$ unitary matrix $u_{jk}$ we have
\begin{equation}
\ketPsiSM=\sum_t\phi_t\ket{\phi_t\,b_t},\quad\text{where}\;
\phi_j\ket{\phi_j}=\sum_{s=1}^n\psi_{s}u_{sj}\ket{p_s}\;\text{and}\;
\ket{b_k}=\sum_s u_{sk}^*\ket{a_s}.
\end{equation}
Let us find $u_{jk}$ such that the first element of this \RhoSensemble is our arbitrary vector, that is, $\ket{\phi_1}=\ket{\xi}$. Then
\begin{equation}
\ket{\phi_1}=\sum_{s=1}^n\gamma_s\ket{p_s}=\frac{1}{\phi_1}\sum_{s=1}^n\psi_{s}u_{s1}\ket{p_s},
\end{equation}
whence $u_{s1}=\phi_{1}\gamma_s/\psi_s$. Unitarity requires  $\sum_s|u_{s1}|^2=1=\phi_1\sum_s|\gamma_s/\psi_s|^2$.  Thus we have
\begin{equation}
  \ket{b_1}=\sum_s u_{s1}^*\ket{a_s},\quad\text{with}\;
  u_{j1}=(\gamma_j/\psi_j)/\sqrt{\textstyle\sum_{s=1}^n|\gamma_s/\psi_s|^2}.  
\end{equation}
Arbitrarily complete the orthonormal set $\set{\ket{b_k}}_{k=1}^{\nM}$, which, by (b), must contain an ancilla of a \RhoSensemble containing \ket{\xi}. ${}^{}$\QED

\renewcommand{\baselinestretch}{1.0}\tiny\normalsize

\section{Overview of the several treatments}
\citet{Schrodinger36} established our parts (a), (d), and (e). 

\citet[p.~173]{Jaynes57} attempted to establish our (c) and (d). He required, for a particular matrix $\mathbf{A}$ and a transformation $\mathbf{T}$, the equality $\mathbf{AA^\dagger}=\mathbf{ATT^\dagger A^\dagger}$, which he incorrectly interpreted to require%
\footnote{%
This equality is satisfied by $\mathbf{TT^\dagger}=\mathbf{1}+\mathbf{X}$ with $\mathbf{AXA^\dagger}=0$.
} %
the unitarity of $\mathbf{T}$. Thus Jaynes failed to establish his (correct) claim of isomorphism between the group of unitary transformations and the group of U-mappings of {\RhoSensemble}s. \citet{Hadjisavvas81}, extending Jaynes, elegantly obtained our (c), (d), and (e) for infinite dimensions. It appears that none of the later authors were aware of either Jaynes' or Hadjisavvas's results.

\citet{Gisin89} established a weak (and unphysical --- see below) version of our (a).

HJW establish our (c) and (d) in their theorem's parts~(b) and (a), respectively, and our (a) in the ``application'' in their Section~3.3. (The term ``\RhoSensemble'' is due to HJW; Jaynes called it an ``array''.) 

The ``GHJW Theorem'' presented by Mermin (1999) is essentially HJW's Section~3.3 (\ie our (a)). Our proofs of the lemma and of part (a) of the theorem are quite similar to the development in the appendix of Mermin (1999). (In personal correspondence, David Mermin gives credit for what I've formalized as the Lemma to a conversation with Chris Fuchs.) 

Our (b) (``every complete variable of \SysM determines exactly one \RhoSensemble'') seems to have its first explicit statement here.

We see that this theorem is primarily \Schrodinger's --- not only by priority, but by having established the major portion, parts (a), (d), and (e). Hadjisavvas and HJW added part (c). It is fairly well-known by the initials ``HJW''; taking all this into account, and letting the ``H'' work a little harder, we arrive at ``\Schrodinger-HJW'' as a priority-recognizing name.

\section{Comments and Discussion}
The U-map (orthogonal-columns matrix transformation) approach of \Schrodinger and HJW, though it successfully generates all {\RhoSensemble}s, has no obvious physical significance. Mermin, by treating the ancillary system \SysM as a real physical system with \kPSM as the actual quantum-mechanical state of \SysSM (and utilizing Fuchs' ``lemma''), was able to obtain an arbitrary \RhoSensemble's ancilla (HJW's Section 3.3) without using the U-maps of HJW's theorem. This makes it possible to express the HJW theorem in terms of unitary transformations between observables in the ancillary system (our (c) and (d)).  Instead of the mathematical formality of the U-maps, we have the physical significance of the correlation of each ancillary variable with a {\RhoSensemble}.

In Gisin's treatment, \kPSM is arbitrarily imposed: The state vector of the joint system is constructed in the proof (as was the \ket{\Psi'} of our proof), not assumed given (as was the \kPSM of our theorem). However, it is necessary that the two systems were prepared \emph{ab initio} in a pure joint state (\ie \kPSM) which reduces to the specified mixture; Gisin's ``steering'' theorem, as presented, is not physical. \Schrodinger and HJW make this clear; Mermin does also, in the body of his paper. However, in his appendix, Mermin states that, for any mixed state of Alice's system, it is always ``possible to provide Bob with a system of his own for which the joint Alice-Bob system has a pure state'' which reduces to Alice's mixture. As expressed, this statement may mislead: For example, if Alice's system is mixed because it is already entangled in a pure joint state with Carol's system, then no such system may be provided to Bob (except by stealing Carol's). That is, it is not possible to introduce the ancillary system \emph{post facto} --- it and the system must be initialized together.

To \Schrodinger, the central point was that ``\emph{in general} a sophisticated experimenter can, by a suitable device which does \emph{not} involve measuring non-commuting variables, produce a non-vanishing probability of driving the system into any state he chooses'' by means of measurement on the entangled ancilla.   Unfortunately, \Schrodinger's phrase ``driving the system into any state he chooses'' tends to overwhelm the modifier thereof, ``a non-vanishing probability,'' so \Schrodinger's point tends to be misstated; for example, \citet[p.~221]{Jammer74} summarizes \Schrodinger's statement ``an experimenter can indeed steer a far-away system, without interfering with it at all, into any state out of an infinity of possible states\dots,'' leaving an incorrect impression of deterministic control of the outcome. 

The \Schrodinger-HJW Theorem gives a complete catalog of potential correlations between the {\RhoSensemble}s of \SysS and the disjoint (orthogonal) sets of states of \SysM, in the case that \SysSM was prepared in a pure state. This catalog allows the design of an experiment involving a measurement on \SysM which makes exactly one \RhoSensemble ``visible'' --- that is, allows every  occurrence of \SysS to be assigned a particular state of the \RhoEnsemble without ``disturbing'' \SysS. To be more precise, the measurement%
\footnote{%
A ``measurement'' of \Op{B} is an interaction of \SysM with \Sys{X} (a system external to \SysSM) such that $\ket{b_j\,\xi}\longrightarrow\ket{b_j\,x_j}$, the \set{\ket{x_j}\in\HSX} orthonormal; the state of \SysSM changes from pure to mixed as a direct result of this unitary transformation on \HSSMX. So long as \SysSM remains isolated, not only do the ancilla and $\bRho$-ensemble remain correlated, but also the \set{\ket{b_j}} maintain their correlation with disjoint elements of the exterior (initially the \set{\ket{x_j}}), regardless of further external interactions with \Sys{X}; von Neumann's infinite regression of measurements is irrelevant to our discussion.
} %
of an ancilla (say, an observable \Op{B} of \SysM with eigenkets \set{\ket{b_j}}, the ancilla of the \RhoSensemble \set{\big(\ket{\phi_j},\,w_j\big)}) changes the state of \SysSM from the pure state $\ketPsiSM=\sum_s\sqrt{w_s}\,e^{i\theta_s}\ket{\phi_s\,b_s}$ to the mixture $\RhoSM=\sum_s w_s\,\proj{\phi_s\,b_s}$. The correlation of this particular ancilla with this particular \RhoSensemble is unique%
\footnote{%
The correlation is unique even in the case of degenerate probabilities (\eg EPR).
} %
\citep{Kirkpatrick02} --- none of the other putative correlations in the \Schrodinger-HJW theorem survive the measurement of the ancilla.

\bigskip\noindent\textbf{Acknowledgments}\medskip\par%
Thanks to Richard Jozsa for pointing out the Jaynes and Hadjisavvas articles, and to David Mermin for generously sharing his recollections regarding the development of the proof he published.

 \renewcommand{\refname}{\normalfont\sc References}
 \footnotesize%


\begin{thebibliography}{}

\bibitem[Gisin(1989)]{Gisin89}
Gisin, N. (1989).
``Stochastic quantum dynamics and relativity,'' \emph{Helvetica Physica Acta}
  {\bf 62}, 363--371.

\bibitem[Hadjisavvas(1981)]{Hadjisavvas81}
Hadjisavvas, N. (1981).
``Properties of mixtures on non-orthogonal states,'' \emph{Lett. Math. Phys.}
  {\bf 5}, 327--332.

\bibitem[Hughston et~al.(1993)Hughston, Jozsa, and Wootters]{HughstonJW93}
Hughston, L.~P., Jozsa, R., and Wootters, W.~K. (1993).
``A complete classification of quantum ensembles having a given density
  matrix,'' \emph{Phys. Lett. A} {\bf 183}, 14--18.

\bibitem[Jammer(1974)]{Jammer74}
Jammer, M. (1974).
{\em The Philosophy of Quantum Mechanics}.
Wiley, New York.

\bibitem[Jaynes(1957)]{Jaynes57}
Jaynes, E.~T. (1957).
``Information theory and statistical mechanics. {II},'' \emph{Phys. Rev.} {\bf
  108}(2), 171--190.

\bibitem[Kirkpatrick(2002)]{Kirkpatrick02}
Kirkpatrick, K.~A. (2002).
``Uniqueness of a convex sum of products of projectors,'' \emph{J. Math. Phys.}
  {\bf 43}(1), 684--686, \eprint{quant-ph/0104093}.

\bibitem[Mermin(1999)]{Mermin99}
Mermin, N.~D. (1999).
``What do these correlations know about reality? {N}onlocality and the
  absurd,'' \emph{Found. Phys.} {\bf 29}(4), 571--587,
  \eprint{quant-ph/9807055}.

\bibitem[Schr{\"o}dinger(1936)]{Schrodinger36}
Schr{\"o}dinger, E. (1936).
``Probability relations between separated systems,'' \emph{Proc. Camb. Phil.
  Soc.} {\bf 32}, 446--452.

\end{thebibliography}
\end{document}